\documentclass{Interspeech}

\interspeechcameraready 

\title{MATER: Multi-level Acoustic and Textual Emotion Representation for Interpretable Speech Emotion Recognition\thanks{$^*$Equal contribution.}}

\author[affiliation={1,*}]{Hyo Jin}{Jon}
\author[affiliation={1,2,*}]{Longbin}{Jin}
\author[affiliation={1}]{Hyuntaek}{Jung}
\author[affiliation={1}]{Hyunseo}{Kim}
\author[affiliation={2}]{Donghun}{Min}
\author[affiliation={1,2}]{Eun Yi}{Kim}

\affiliation{Artificial Intelligence \& Computer Vision Lab.}{Konkuk University}{South Korea}
\affiliation{AI Engine Development Team}{Voinosis Inc.}{South Korea}
\email{\{hyojin2011, busan199, hs11015, eykim\}@konkuk.ac.kr, \{lbjin, dhmin\}@voinosis.com}

\usepackage{comment}
\usepackage{algorithm}
\usepackage{algorithmicx}
\usepackage{algpseudocode}
\usepackage{booktabs}
\usepackage{multirow}
\usepackage{listings}
\usepackage{arydshln}
\usepackage{array}
\newcolumntype{L}[1]{>{\raggedright\let\newline\\\arraybackslash\hspace{0pt}}m{#1}}
\newcolumntype{C}[1]{>{\centering\let\newline\\\arraybackslash\hspace{0pt}}m{#1}}
\newcolumntype{R}[1]{>{\raggedleft\let\newline\\\arraybackslash\hspace{0pt}}m{#1}}

\begin{document}

\maketitle

\begin{abstract}
This paper presents our contributions to the Speech Emotion Recognition in Naturalistic Conditions (SERNC) Challenge, where we address categorical emotion recognition and emotional attribute prediction. To handle the complexities of natural speech, including intra- and inter-subject variability, we propose Multi-level Acoustic-Textual Emotion Representation (MATER)—a novel hierarchical framework that integrates acoustic and textual features at the word, utterance, and embedding levels. By fusing low-level lexical and acoustic cues with high-level contextualized representations, MATER effectively captures both fine-grained prosodic variations and semantic nuances. Additionally, we introduce an uncertainty-aware ensemble strategy to mitigate annotator inconsistencies, improving robustness in ambiguous emotional expressions. MATER ranks fourth in both tasks with a Macro-F1 of 41.01\% and an average CCC of 0.5928, securing second place in valence prediction with an impressive CCC of 0.6941.
\end{abstract}

\section{Introduction}
Emotions are fundamental to human experience, influencing decision-making, communication, and mental well-being \cite{kakunje2020emotional}. Speech, as the most common form of communication, conveys a rich blend of emotional cues, including speaker identity, affect, and linguistic emphasis \cite{picard2000affective}. Consequently, speech emotion recognition (SER) has gained significant attention over the past two decades for its ability to recognize emotions across categorical labels \cite{ekman1992argument} or continuous emotional dimensions \cite{russellnmehrabian}.

Despite extensive research, most SER datasets fail to fully capture real-world emotional expressions, as they are often composed of acted \cite{burkhardt2005database, grimm2008vera, jackson2014surrey} or elicited \cite{iemocap, crema-d} recordings.
Even datasets featuring natural speech \cite{busso2016msp, li2017cheavd, luo2018investigation} are often constrained by a limited number of participants, reducing their generalizability to diverse real-world scenarios. To address these issues, the MSP-Podcast corpus \cite{lotfian2017_msp_podcast_dataset}, provided in the Speech Emotion Recognition in Naturalistic Conditions (SERNC) Challenge \cite{interspeech_ser_challenge}, offers natural speech from 2,826 speakers, supporting two tasks: Task 1 (\textit{Categorical Emotion Recognition}) and Task 2 (\textit{Emotional Attribute Prediction}).

\begin{figure}[t]
  \centering
  \includegraphics[width=\linewidth]{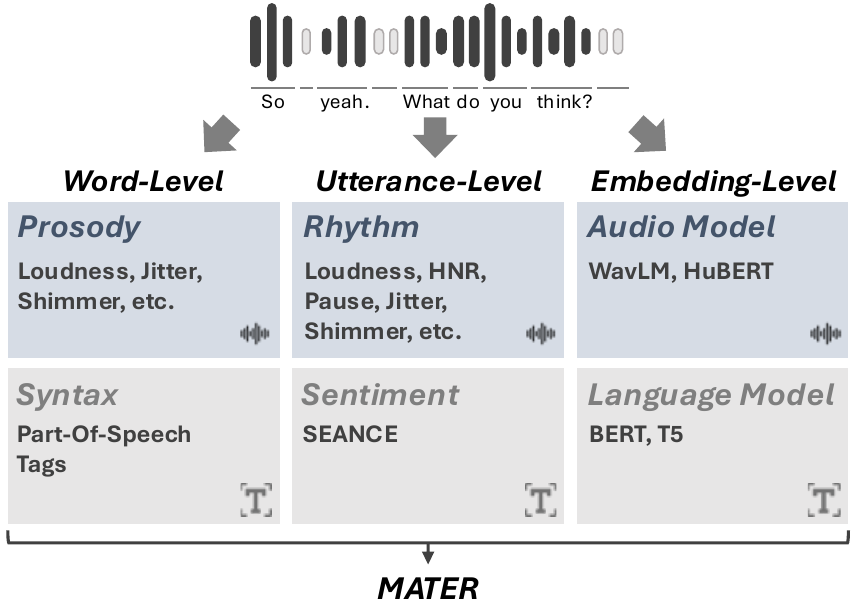}
  \caption{The MATER framework that integrates word-, utterance-, and embedding-level features to construct robust and context-aware emotion representations.}
  \label{fig:MATER}
\end{figure}

One of the primary challenges in naturalistic SER is developing representations that are robust to intra- and inter-speaker variability while effectively capturing the complexity of emotional expression. To address this, we introduce \textit{Multi-level Acoustic and Textual Emotion Representation} (MATER)—a novel hierarchical framework that integrates acoustic and textual features across multiple levels for a comprehensive understanding of emotions. As illustrated in Figure \ref{fig:MATER}, MATER systematically models emotions at three distinct levels: 1)	Word-level: Extracting syntax-aware prosodic cues to capture the fine-grained interplay between linguistic structure and intonation. 2)	Utterance-level: Analyzing sentiment and rhythmic patterns over the entire speech segment to capture both lexical meaning and overall emotional nuances. 3) Embedding-level: Leveraging pretrained deep models (\textit{e.g., WavLM, HuBERT}) to derive context-sensitive and speaker-invariant representations for better generalization. This multi-level framework enables MATER to model emotions from low-level syntactic and prosodic cues to high-level contextualized representations, effectively integrating both speech and text for robust emotion modeling.

A further challenge in SER with natural speech arises from overlapping, ambiguous, and highly variable emotional expressions, which often hinder annotator consensus. This inconsistency introduces confidence disparities and biases across emotion categories. To tackle this, we propose a novel uncertainty-aware ensemble that improves robustness by selecting emotion predictions with the least uncertainty, thereby mitigating annotation biases and enhancing reliability.

MATER achieves high-ranking performance in both tasks of the SERNC Challenge, with a Macro-F1 of 41.01$\%$ and an average CCC of 0.5928. The results underscore its effectiveness in capturing fine-grained emotional nuances in natural speech.

\begin{figure}[t]
  \centering
  \includegraphics[width=0.95\linewidth]{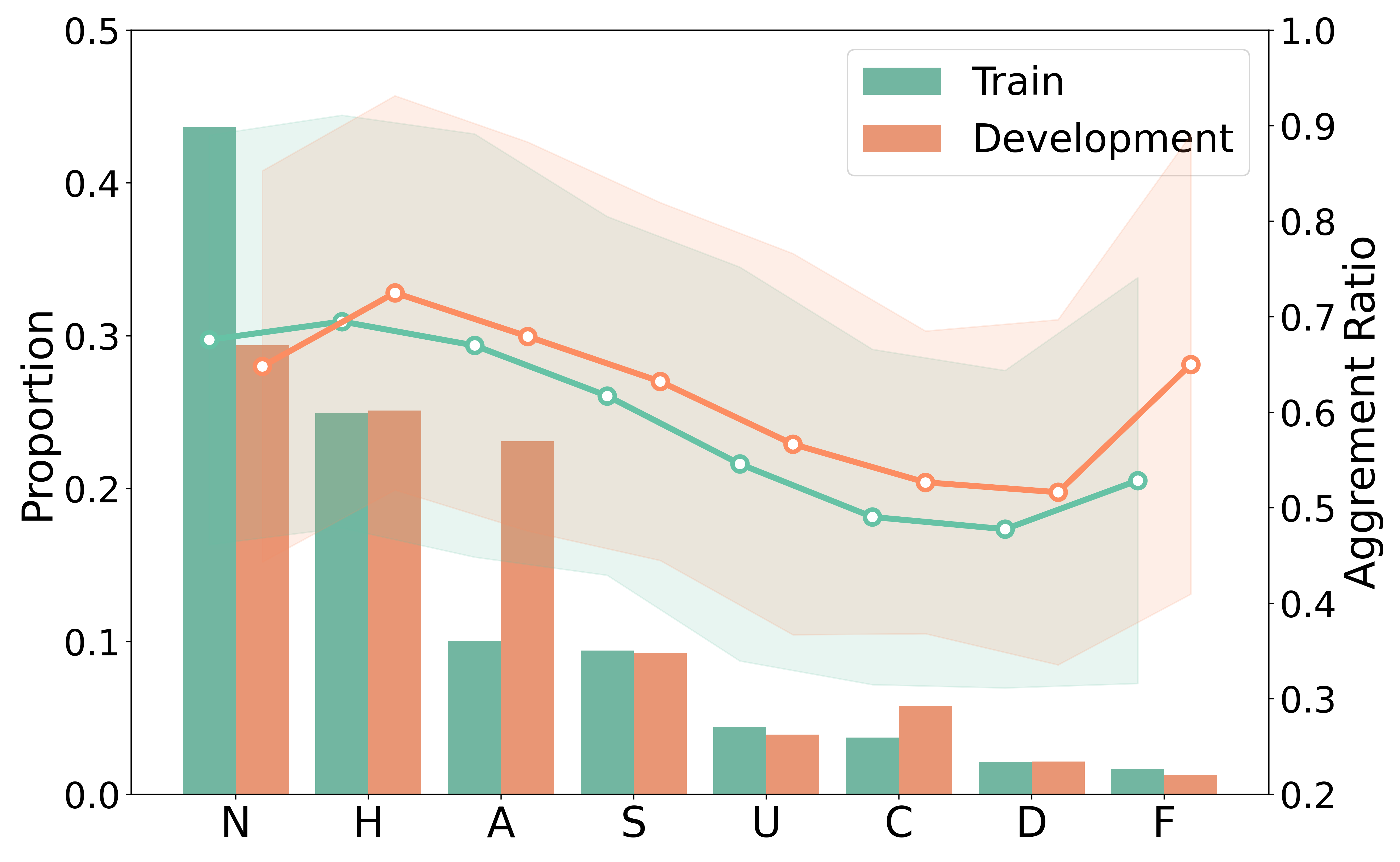}
  \caption{Emotional category distribution and annotator agreement in the train and development sets for Task 1. Bars represent sample proportions, highlighting class imbalance, while lines indicate average agreement ratios. Shaded areas denote standard deviation, illustrating consensus inconsistencies.}
  \label{fig:data}
\end{figure}

\section{Dataset}
This study utilizes the MSP-Podcast corpus \cite{lotfian2017_msp_podcast_dataset}, provided by the organizers of the SERNC Challenge. This dataset contains short naturalistic speech segments extracted from podcasts. The dataset is annotated for two tasks:
\begin{enumerate}
    \item \textit{Categorical Emotion Recognition}—Speech segments are labeled into eight emotion categories; Angry (A), Contempt (C), Disgust (D), Fear (F), Happy (H), Neutral (N), Sad (S), and Surprise (U). Samples labeled as “Other” or those lacking annotator agreement are excluded.
    \item \textit{Emotional Attribute Prediction}—Speech is rated on a seven-point Likert scale along three emotional dimensions: arousal (calm to active), dominance (weak to strong), and valence (negative to positive).
\end{enumerate}

The train and development sets contain 84,260 samples from 2,112 speakers and 31,961 samples from 714 speakers, respectively. The test set comprises 3,200 balanced samples across eight emotion categories. 
However, unlike the train and development sets, it does not provide transcripts or additional metadata. To ensure consistency, we generate transcripts and forced alignments for all sets using \textit{Whisper-large-v3} \cite{whisper}.

Given its real-world nature, the dataset presents challenges such as class imbalance and inconsistent annotation consensus across emotion categories, as illustrated in Figure \ref{fig:data}. To address the class imbalance, we evaluate models using the average performance across five in-house test sets, each randomly sampled from the development set with 326 samples per emotion. Additionally, to manage annotation uncertainties, we introduce a novel ensemble strategy that employs a non-parametric approach to enhance prediction reliability.

\section{MATER}

\subsection{Feature Extraction}
To model emotional expressions in natural speech, we extract acoustic and textual features across three different levels. Each level captures complementary information, enabling the model to learn both fine-grained speech patterns and high-level contextual cues. The following sections detail each feature level.

\noindent
\textbf{Word-level} 
Word-level features encode both syntactic and prosodic aspects of speech. A BERTweet-based syntactic parser \cite{postag} extracts linguistic patterns, including grammatical person information for pronouns, forming a 20-dimensional syntactic feature vector per word. To capture prosodic variations, a 22-dimensional feature vector is extracted using the openSMILE library \cite{opensmile}, incorporating attributes like loudness, jitter, shimmer, alpha ratio, and voiced/unvoiced segment statistics. By concatenating these features, MATER creates a syntax-aware prosodic representation, allowing the model to learn the interplay between linguistic structure and intonation, which plays a crucial role in emotion expression.

\begin{figure}[t]
  \centering
  \includegraphics[width=0.9\linewidth]{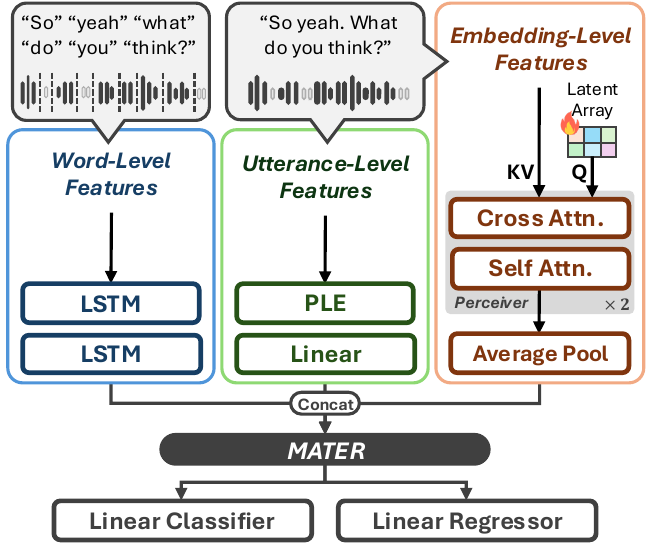}
  \caption{Feature aggregation of MATER. Embedding-level features are aggregated through the Perceiver, utterance-level features use PLE with a linear layer, and word-level features use a two-layer LSTM. The concatenated features are passed to a linear classifier for Task 1 and a linear regressor for Task 2.}
  \label{fig:method}
\end{figure}

\noindent
\textbf{Utterance-level} 
Utterance-level features provide a broader perspective on the emotional characteristics of speech by modeling sentiment and rhythmic patterns. Sentiment features are derived from the SEANCE feature set \cite{seance}, producing a 517-dimensional representation that encapsulates affective tendencies across the entire transcript. Rhythmic features, on the other hand, are extracted to analyze the flow, intensity, and nuances of speech. These include loudness, jitter, shimmer, Harmonic-to-Noise Ratio (HNR), pauses, and voiced/unvoiced segment statistics, resulting in a 34-dimensional feature vector. By incorporating both sentiment and rhythm, the model captures not only the meaning of words but also how they are delivered, which is critical for distinguishing subtle emotional variations.

\noindent
\textbf{Embedding-level} 
Embedding-level features leverage pretrained audio and text encoders to derive context-sensitive and speaker-invariant representations. The audio encoders, \textit{WavLM} \cite{wavlm} and \textit{HuBERT} \cite{hubert}, capture rich phonetic and prosodic information, while the text encoders, \textit{BERT} \cite{bert} and \textit{T5} \cite{t5}, provide semantically informed representations of spoken content. To enhance domain adaptation, we post-pretrain these encoders on the MSP-Podcast corpus, following the baseline’s strategy of fine-tuning \textit{WavLM} with attentive statistics pooling \cite{attentivestatpool}. Empirical results confirm that models trained with domain-adapted embeddings improve SER performance.

\begin{table*}[t!]
  \caption{Performance on categorical emotion recognition. Macro-F1($\%$) and Accuracy($\%$) are reported for the development set, in-house test sets, and the official test set. Final submission results are bolded.}
  \label{tab:task1}
  \centering
  \begin{tabular}{lccccccc}
    \toprule
    \multicolumn{1}{c}{\multirow{2}{*}{\textbf{Feature}}} & \multirow{2}{*}{\textbf{Method}} & \multicolumn{2}{c}{\textbf{Development Set}} & \multicolumn{2}{c}{\textbf{In-house Test Sets}} & \multicolumn{2}{c}{\textbf{Test Set}}\\
    \cmidrule(lr){3-4}\cmidrule(lr){5-6}\cmidrule(lr){7-8}
     &  & \textbf{Macro-F1} & \textbf{Accuracy} & \textbf{Macro-F1} & \textbf{Accuracy} & \textbf{Macro-F1} & \textbf{Accuracy} \\
    \midrule
    \textit{WavLM} (\textit{Baseline}) & \multirow{2}{*}{Fine-tuning} & 31.70 & 44.90 & 35.99 & 39.46 & 32.93 & 35.56 \\
    $\quad\quad\ \ \ $+ Soft Labels &  & 35.43 & 46.54 & 41.55 & 42.59 & - & - \\
    \hdashline[0.5pt/1pt]
    $M_1.$ \textit{HuBERT} & \multirow{2}{*}{Fine-tuning} & 35.81 & 51.98 & 34.71 & 37.66 & - & - \\
    $M_2. \quad$+ Soft Labels &  & 35.91 & 47.05 & 41.77 & 42.91 & - & - \\
    \hdashline[0.5pt/1pt]
    $M_3.$ \textit{HuBERT} + \textit{BERT} & \multirow{3}{*}{MATER} & 37.08 & 48.82 & 40.71 & 40.94 & - & - \\
    $M_4. \quad$+ Soft Labels &  & 37.20 & 48.88 & 40.91 & 41.11 & - & - \\
    $M_5. \qquad$+ Utterance + Word &  & 37.13 & 48.64 & 41.11 & 41.34 & - & - \\
    \hdashline[0.5pt/1pt]
    $M_6.$ \textit{HuBERT} + \textit{T5} & \multirow{5}{*}{MATER} & 37.17 & 48.52 & 42.24 & 42.39 & - & - \\
    $M_7. \quad$+ Soft Labels &  & 37.29 & 47.60 & 42.80 & 42.71 & - & - \\
    $M_8. \qquad$+ Utterance &  & 37.40 & 48.52 & 43.04 & 43.11 & - & - \\
    $M_9. \quad\quad$+ Word &  & 37.48 & 48.43 & 43.45 & 44.72 & - & - \\
    $M_{10}. \, \, \ \quad$+ Utterance + Word &  & 37.61 & 48.43 & 43.49 & 43.72 & - & - \\
     \midrule
    \multicolumn{2}{l}{Averaging Ensemble ($M_2, M_5, M_{10}$)} & - & - & 44.24 & 44.32 & - & - \\
    \multicolumn{2}{l}{Majority Ensemble ($M_2, M_5, M_{10}$)} & - & - & 44.74 & 44.64 & - & - \\
    \multicolumn{2}{l}{Uncertainty-Aware Ensemble ($M_2, M_5, M_{10}$)} & - & - & 44.89 & 45.13 & \textbf{41.01} & \textbf{40.97} \\
    \bottomrule
  \end{tabular}
\end{table*}

\subsection{Feature Aggregation}
MATER integrates multi-level features through dedicated processing architectures. Figure \ref{fig:method} illustrates the overall aggregation process. At the word-level, features are fed into a two-layer LSTM \cite{lstm}, where the final hidden state serves as the word-level embedding, effectively capturing sequential dependencies. At the utterance-level, features are first processed through a piecewise linear embedding (PLE) layer \cite{pleembeddings}, followed by a linear layer to produce a fixed-dimensional representation. At the embedding-level, when multiple embedding sources are used, a Perceiver architecture \cite{perceiver} fuses them into a single representation. Otherwise, the pooled features are used directly without additional processing. Finally, the concatenated multi-level embeddings are fed into linear layers to predict emotion categories or continuous emotional attributes, providing a comprehensive and context-aware emotion representation.

\begin{algorithm}[t]
\caption{Pseudo-code for Uncertainty-Aware Ensemble}
\label{pseudocode}
\begin{algorithmic}
\State{$C\, =$ \{A, C, D, S, H, U, F, N\} : Emotion categories \\
 $M = \{M_1, M_2, ..., M_k\}$ : Candidate models \\
$P_{i} = \{\{p^{i}_{j, c}\}_{j=1}^m\}_{c\in C}$ : Predicted probs. from model $M_i$ \\
$U_{i} = \{\{u^{i}_{j, c}\}_{j=1}^m\}_{c\in C}$ : Uncertainties from model $M_i$ \\
$\hat{y}\in\mathbb{R}^m$ : Ensembled  prediction for $m$ inputs} 
\State
\vspace{-1ex}
\State{$\#$ 1. Select top models from $n$ feature combinations.}
\State{$M' \leftarrow$ Choose top-performing models from $M$}
\State{$\#$ 2. Rank probabilities independently for each model and emotion category from $M'$.}
\For{$i=1$ to $n$}{
\For{$c \in C$}{
\\
$\quad\quad\quad \{u^{i}_{j, c}\}_{j=1}^m \leftarrow $ RANK$(\{p_{j, c}^{i}\}_{j=1}^m)$
}\EndFor
}
\EndFor
\State{$\#$ 3. Assign the emotion with minimal average uncertainty.}
\For{$j=1$ to $m$}{
\\
$\ \, \,\quad \hat{y}_j=\underset{c}{argmin}(\{\frac{1}{n}\sum_{i=1}^nu^{i}_{j, c}\}_{c\in C})$
}
\EndFor \\

\Return{$\hat{y}$}
\end{algorithmic}
\end{algorithm}

\section{Uncertainty-Aware Ensemble}
SER faces inherent challenges due to intra- and inter-speaker variability, as well as overlapping and ambiguous emotional expressions. These factors contribute to annotation inconsistencies, leading to biased predictions and reduced model reliability. Certain emotions, such as contempt and disgust, exhibit lower annotator consensus, resulting in unstable confidence scores and high misclassification errors. Previous attempts, such as Shamsi et al. \cite{shamsi2024conilium}, sought to mitigate these biases by optimizing decision thresholds, but such approaches often lead to overfitting and fail to generalize across diverse evaluation sets.

To address this issue, we introduce a novel uncertainty-aware ensemble strategy designed for balanced evaluation sets. Instead of tuning the decision thresholds, we estimate epistemic uncertainty by ranking predicted probabilities within each emotion category. The final ensemble prediction is obtained by averaging these rank-based uncertainties from the top-performing models across feature combinations, prioritizing high-confidence predictions. This approach reduces ambiguity and ensures robust predictions across emotion categories, even with biased label consensus. Additionally, unlike parametric methods, our rank-based method requires no additional optimization, making it inherently adaptable and generalizable across datasets. These advantages provide a scalable and reliable solution for improving SER in real-world settings. The detailed ensemble process is outlined in Algorithm \ref{pseudocode}.

\begin{table*}[t!]
  \caption{Performance on emotional attribute prediction. CCC is reported for valence, arousal, and dominance on the development and test sets, with final submission results in bold.}
  \label{tab:task2}
  \centering
  \begin{tabular}{L{3.5cm}C{1.75cm}C{1.75cm}C{1.75cm}C{1.75cm}C{1.75cm}C{1.75cm}}
    \toprule
    \multicolumn{1}{c}{\multirow{2}{*}{\textbf{Feature}}} & \multicolumn{3}{c}{\textbf{Development Set}} & \multicolumn{3}{c}{\textbf{Test Set}} \\
    \cmidrule(lr){2-4}\cmidrule(lr){5-7}
     & \textbf{Valence} & \textbf{Arousal} & \textbf{Dominance} & \textbf{Valence} & \textbf{Arousal} & \textbf{Dominance} \\
    \midrule
    \textit{WavLM} (\textit{Baseline}) & .7345 & .6925 & .6246 & .6385 & .6232 & .4775 \\
    $\quad\quad\ \ \ $+ Utterance + Word & .7381 & .6901 & .6264 & - & - & \textbf{.4722} \\
    \hdashline[0.5pt/1pt]
    \textit{WavLM} + \textit{T5} & .7636 & .6948 & .6289 & - & - & - \\
    $\quad\quad\ \ \ $+ Utterance & .7621 & .7043 & .6347 & - & \textbf{.6119} & .4552 \\
    $\quad\quad\ \ \ $+ Utterance + Word & .7733 & .6885 & .6192 & \textbf{.6941} & - & -  \\
    \bottomrule
  \end{tabular}
\end{table*}

\section{Experiments}

\subsection{Experimental Setup}
We evaluate MATER and its individual feature levels. For individual levels, word-level features are processed using a two-layer LSTM (128 hidden units), while utterance-level features are classified via SVM with an RBF kernel. Embedding-level features are fine-tuned using \textit{large} models with attentive statistical pooling, following the baseline. In MATER, word- and utterance-level features are projected into 128-dimensional vectors, while the Perceiver produces a 768-dimensional output with a $64\times768$ latent array. Each embedding-level feature passes through the Perceiver block twice before integration.

Loss functions are task-specific: weighted cross-entropy for Task 1 and concordance correlation coefficient (CCC) loss for Task 2. Models are trained for 50 epochs with learning rates from $1\times10^{-5}$ to $5\times10^{-7}$ and batch sizes between 128 and 2048. Evaluation follows the SERNC Challenge, using Macro-F1 for Task 1 and CCC for Task 2.

\subsection{Performance on Categorical Emotion Recognition}
MATER achieves progressive improvements across feature levels, validating the effectiveness of its multi-level fusion. As shown in Table \ref{tab:task1}, word-level features contribute more significantly than utterance-level features, suggesting that syntax-aware prosody is more informative for categorical emotion recognition. While soft labels are effective in fine-tuned models, they provide only marginal benefits in MATER. A possible reason is that full fine-tuning causes overfitting to rigid label boundaries, whereas MATER mitigates this issue by preserving pretrained encoder representations. Among ensemble strategies, the uncertainty-aware ensemble surpasses averaging and majority voting, reinforcing the importance of uncertainty estimation in emotion classification. This highlights how ranking-based uncertainty estimation mitigates annotation biases and improves decision reliability. Ultimately, MATER achieves a Macro-F1 of 41.01$\%$, attaining fourth place in Task 1.

\subsection{Performance on Emotional Attribute Prediction}
MATER effectively captures emotional polarity through its multi-level integration of acoustic and textual features. As shown in Table \ref{tab:task2}, valence prediction improves significantly, whereas gains in arousal and dominance remain limited. This aligns with the intuition that valence is more text-dependent, while arousal and dominance are driven by acoustic cues. Since MATER integrates both modalities, its text-heavy fusion may not fully exploit arousal- and dominance-related acoustic variations. Despite this, MATER remains competitive across all emotional dimensions, achieving a strong overall ranking.

\begin{figure}[t]
  \centering
  \includegraphics[width=\linewidth]{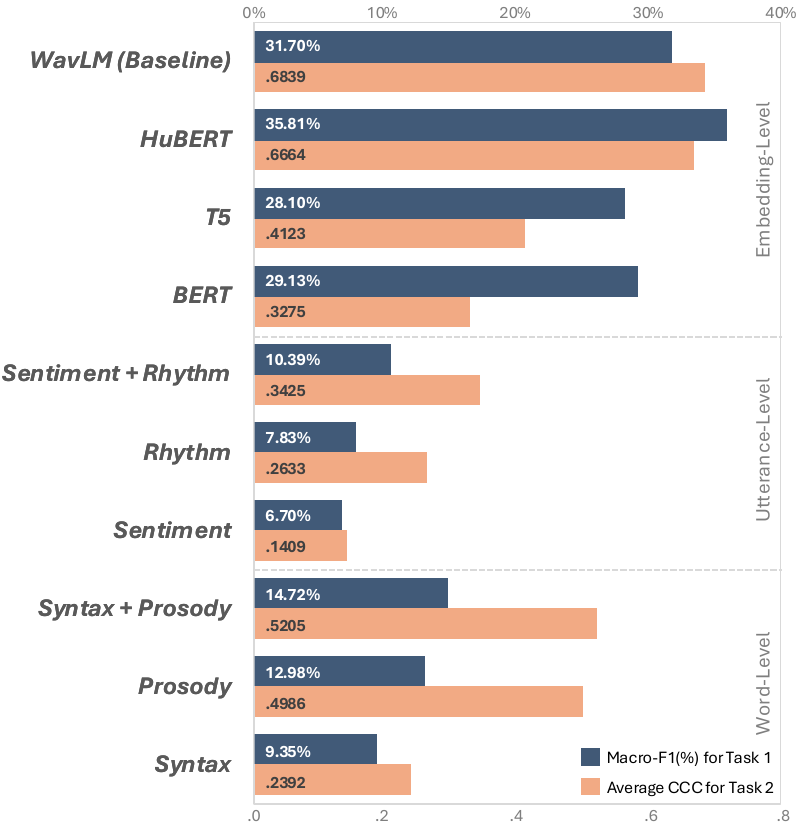}
  \caption{Performance comparison across feature levels for Task 1 (Macro-F1) and Task 2 (average CCC for valence, arousal, and dominance) on the development set.}
  \label{fig:feature_levels}
\end{figure}

\subsection{Post-Challenge Analysis on MATER}
Due to time constraints during the challenge, a detailed feature impact analysis was not fully conducted. Post-challenge investigations reveal key correlations between feature sets and task performance, offering interpretable insights for further refinement.

As shown in Figure \ref{fig:feature_levels}, acoustic features consistently outperform textual ones across both tasks, highlighting the importance of prosodic and spectral variations in SER. At the embedding-level, optimal encoders differ across tasks, emphasizing the need for task-specific encoder selection. Additionally, multimodal fusion in MATER enhances performance at the word and utterance levels, confirming its effectiveness in integrating acoustic and textual cues. These findings suggest that adaptive feature weighting or task-driven feature selection could further optimize MATER’s performance. Given that arousal and dominance rely more on acoustic cues, a dynamic fusion strategy balancing audio-text integration may yield additional gains.

\section{Conclusion}
We introduce MATER, a hierarchical framework that integrates acoustic and textual features across word, utterance, and embedding levels, capturing both low-level cues and high-level contextual representations. Our uncertainty-aware ensemble enhances robustness against annotation inconsistencies, improving reliability in ambiguous emotional expressions. MATER ranked fourth in both tasks of the SERNC Challenge, achieving a Macro-F1 of 41.01\% and an average CCC of 0.5928, notably securing the second place in valence prediction. Despite its effectiveness, challenges remain in arousal and dominance prediction. Future work will focus on emotion-specific feature selection and extending MATER to diverse SER datasets, advancing toward a more robust real-world SER system. 

\section{Acknowledgement}
We would like to thank the members of Artificial Intelligence \& Computer Vision Lab. at Konkuk university, including Heejae Choi, Shinwoo Ham, and Junil Jeon, as well as Seokjin Lee from Voinosis Inc., for their valuable support during the SERNC Challenge. Their dedication and assistance contributed greatly to this achievement.

This work was supported by Institute of Information \& communications Technology Planning \& Evaluation (IITP) under the metaverse support program to nurture the best talents (IITP-2025-RS-2023-00256615) grant funded by the Korea government(MSIT).

\bibliographystyle{IEEEtran}
\bibliography{mybib}

\end{document}